\documentclass[conference]{IEEEtran}
\IEEEoverridecommandlockouts
\usepackage{cite}
\usepackage{amsmath,amssymb,amsfonts}
\usepackage{algorithmic}
\usepackage{graphicx}
\usepackage{textcomp}
\usepackage{xcolor}
\usepackage{blindtext}
\usepackage{fvextra}
\usepackage{multirow}
\usepackage[normalem]{ulem}
\usepackage{here}
\usepackage[subrefformat=parens]{subcaption}
\usepackage{booktabs}
\usepackage{hyperref}

\def\BibTeX{{\rm B\kern-.05em{\sc i\kern-.025em b}\kern-.08em
    T\kern-.1667em\lower.7ex\hbox{E}\kern-.125emX}}
\begin{document}

\title{Rule-Based Error Classification for Analyzing Differences in Frequent Errors
\thanks{This work was supported by the Japan Society for the Promotion of Science (JSPS) KAKENHI Grant Number JP23H03508.}
}

\author{
\IEEEauthorblockN{Atsushi Shirafuji}
\IEEEauthorblockA{
\textit{University of Aizu}\\
Aizu-Wakamatsu, Japan \\
m5261161@u-aizu.ac.jp
}
\and
\IEEEauthorblockN{Taku Matsumoto}
\IEEEauthorblockA{
\textit{Japan Atomic Energy Agency}\\
Naraha, Futaba, Japan \\
matsumoto.taku27@jaea.go.jp
}
\and
\IEEEauthorblockN{Md Faizul Ibne Amin}
\IEEEauthorblockA{
\textit{University of Aizu}\\
Aizu-Wakamatsu, Japan \\
aminfaizul007@gmail.com
}
\and
\IEEEauthorblockN{Yutaka Watanobe}
\IEEEauthorblockA{
\textit{University of Aizu}\\
Aizu-Wakamatsu, Japan \\
yutaka@u-aizu.ac.jp
}
}

\maketitle

\begin{abstract}
Finding and fixing errors is a time-consuming task not only for novice programmers but also for expert programmers.
Prior work has identified frequent error patterns among various levels of programmers.
However, the differences in the tendencies between novices and experts have yet to be revealed.
From the knowledge of the frequent errors in each level of programmers, instructors will be able to provide helpful advice for each level of learners.

In this paper, we propose a rule-based error classification tool to classify errors in code pairs consisting of wrong and correct programs.
We classify errors for 95,631 code pairs and identify 3.47 errors on average, which are submitted by various levels of programmers on an online judge system.
The classified errors are used to analyze the differences in frequent errors between novice and expert programmers. 

The analyzed results show that, as for the same introductory problems, errors made by novices are due to the lack of knowledge in programming, and the mistakes are considered an essential part of the learning process.
On the other hand, errors made by experts are due to misunderstandings caused by the carelessness of reading problems or the challenges of solving problems differently than usual.

The proposed tool can be used to create error-labeled datasets and for further code-related educational research.
\end{abstract}


\section{Introduction}
\label{sec:introduction}

Programming education has been mandatory for primary education in Japan since 2020, but educational inequality has become an issue.
Some students have difficulty accessing proper education due to problems in the setup of the environment and a lack of instructors, especially in rural areas.
Instructors also struggle to prepare lecture materials and provide proper advice because of the variation of knowledge and experience among learners.

Using an online judge (OJ) system has become increasingly popular as an online education platform~\cite{wasik2018oj}.
It allows learners in remote areas to learn programming using the same platform without the setup of the environment.
It is used by various levels of programmers; novices learn programming, and experts test their skills.
Users are asked to write a program to solve the chosen programming problem in a browser.
They can virtually test the input and output cases without calling any commands on their environment.
It does not require additional knowledge and allows the users to concentrate on learning the programming language.
The OJ system automatically judges the users' submitted program and returns the verdict.
The verdict is not only the binary correctness; it provides more detailed failure reasons\footnote{\url{https://onlinejudge.u-aizu.ac.jp/judges_replies}.}, e.g., 
\textit{Time Limit Exceeded}, or \textit{Memory Limit Exceeded}.
They often submit the wrong program (WA) that contains errors\footnote{Several terms are often used synonymously: \textit{errors}, \textit{bugs}, \textit{defects}, \textit{faults}, and \textit{vulnerabilities}. We use \textit{errors} in this paper.}.
They fix the errors in the WA and re-submit it as an accepted program (AC).

One of the popular OJ systems, Aizu Online Judge (AOJ)~\cite{watanobe2004aoj}, has approximately 8 million programs submitted by 100 thousand users for 3,000 programming problems~\cite{watanobe2022aoj}.
It also provides pre-note and post-note, which are a commentary of the problem.
The pre-note introduces the knowledge required to solve the problem for struggling learners.
The post-note presents an example solution program and explains the standard approach to solving the problem.
The problems and the submissions on AOJ are often used for research purposes~\cite{rahman2021repair, matsumoto2021repair, terada2021completion, muepu2023recommend, kawabayashi2021classify, mostafiz2021classify} and contained in popular datasets such as CodeNet~\cite{puri2021codenet} and CodeContests~\cite{li2022alphacode}.

It is helpful for learners and instructors to know what kinds of programmers make what kinds of errors frequently.
Many prior works examined frequent or common error patterns~\cite{alzahrani2021common, osman2014classify, campos2017common, denny2012classify, kawabayashi2021classify, mostafiz2021classify, ahmadzadeh2005error, ettles2018common, mcall2014meaningful, hristova2003identify, alzahrani2019manual, li2021error, jackson2005error}.
However, the differences in the tendencies between the levels of programmers have yet to be revealed.
Once the differences in the tendencies have been revealed, learners will be able to pay attention to the frequent errors, and they will be able to prevent making similar mistakes.
In addition, instructors will also be able to provide helpful advice about each problem for each learner.

In this work, we analyze the differences in frequent errors between novice and expert programmers.
We propose a rule-based error classification tool that can classify errors based on predefined rules in code pairs consisting of WA and AC submitted by various levels of programmers on AOJ.
We analyze the error classification results based on the user levels.

The main contributions of our work are as follows.
\begin{itemize}
    \item We propose a rule-based error classification tool that can classify both syntax and logic errors.
    \item We show an analysis of the characteristic differences between novice and expert programmers.
    \item We construct an error-labeled dataset containing over 95 thousand code pairs from 44 introductory programming problems.
\end{itemize}
The tool and the dataset can be used for further code-related educational research, such as detecting potential errors or suggesting fixes for incorrect programs.

The rest of this paper is organized as follows.
Section~\ref{sec:related-work} mentions related work in this field.
Section~\ref{sec:method} introduces the method of classification and analysis.
Section~\ref{sec:results} presents and discusses the obtained experimental results.
Section~\ref{sec:limitation} describes the limitations of this research.
Finally, Section~\ref{sec:conclusion} concludes this paper.

\section{Related Work}
\label{sec:related-work}

There have been several approaches for detecting and classifying errors made by programmers~\cite{osman2014classify, campos2017common, denny2012classify, kawabayashi2021classify, mostafiz2021classify, ahmadzadeh2005error, ettles2018common, mcall2014meaningful, hristova2003identify, alzahrani2019manual, li2021error, jackson2005error}.
In particular, to promote programming education by revealing the common and frequent errors made by novice programmers, classifying and analyzing errors for programming education has been performed for decades~\cite{denny2012classify, kawabayashi2021classify, mostafiz2021classify, ahmadzadeh2005error, ettles2018common, mcall2014meaningful, hristova2003identify, alzahrani2019manual, li2021error, jackson2005error}.

Denny~et~al.~\cite{denny2012classify} analyzed syntax errors made by students and found common errors.
They identified that the most common syntax errors are \textit{cannot resolve identifier}, \textit{type mismatch}, and \textit{missing semicolon}.
In addition, the \textit{cannot resolve identifier} and \textit{type mismatch} are reported as the most time-consuming errors; students spent 11.9 minutes and 10.0 minutes resolving the errors on average, respectively.
Osman~et~al.~\cite{osman2014classify} found frequent bug-fix patterns in code change histories from 717 open-source projects on GitHub written in Java.
To capture more essential parts of the code changes, the source code is anonymized and normalized.
They discovered that more than 70\% of bug fixes are less than four lines of code, and about 40\% of the bug-fixing code changes are recurrent.
They reported that the most frequent patterns are \textit{missing null checks}, \textit{missing invocation}, \textit{wrong name}, and \textit{undue invocation}.
Kawabayashi~et~al.~\cite{kawabayashi2021classify} used K-means clustering to find frequent error patterns from code pairs consisting of WA and AC programs written in C++ on AOJ.
For example, they identified that, in a problem that requires reversing the given string and output in a line, the most frequent error is \textit{missing line feeds}.

While automated tools to classify errors have been developed, human analysis is still reliable for some purposes.
Alzahrani~et~al.~\cite{alzahrani2019manual} manually analyzed incorrect programs submitted by students and found common coding errors that cause struggle.

Recently, error classification using machine learning and deep learning has been examined.
Li~et~al.~\cite{li2021error} developed a compilation error classification model using TextCNN~\cite{kim2014textcnn}.

Different from the mentioned prior work, our proposed rule-based tool particularly aims to cover any syntax and logic errors in Python programs by defining comprehensive rules.

\section{Method}
\label{sec:method}

The main phases to analyze the differences in frequent errors are divided into the following: (1) data collection, (2) code normalization, (3) changes extraction, (4) error classification, and (5) difference analysis.

Several prior works used compiler error messages to analyze errors~\cite{jackson2005error, ahmadzadeh2005error, denny2012classify, li2021error}.
However, the error message analysis only focused on syntax errors, which fail in compilation and output error messages.
In this work, we aim to analyze not only syntax errors but also logic errors, which work incorrectly without crashing (e.g., incorrect algorithm or output format).
To identify the cause of these logic errors, we directly analyze the source code rather than the error messages.

\subsection{Data Collection}
\label{sec:method:data-collection}

\begin{figure}[h]
    \centerline{\includegraphics[width=0.6\linewidth]{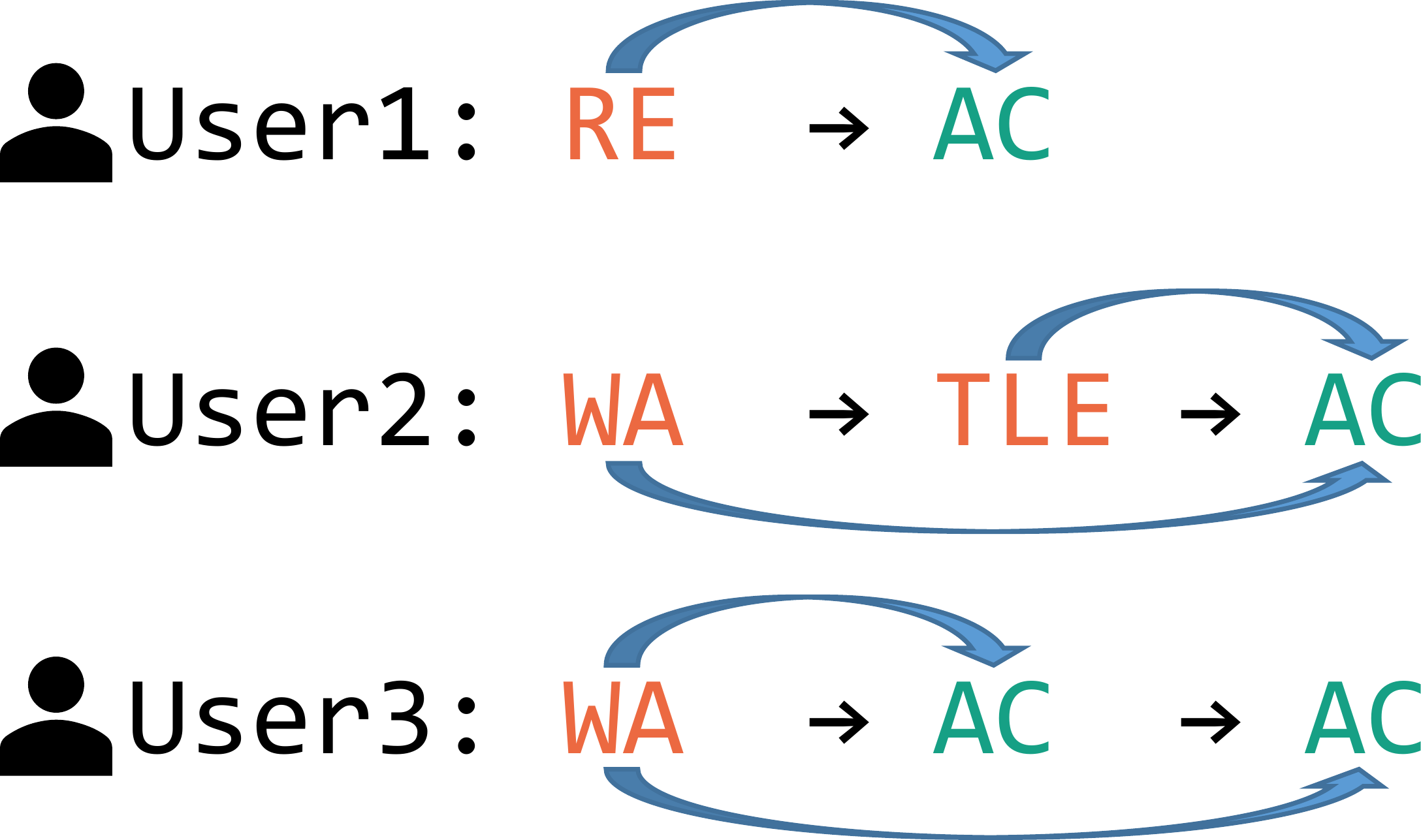}}
    \caption{Illustration of collecting code pairs consisting of wrong and correct programs from the same user. \textit{AC} indicates accepted (correct) programs. \textit{WA}, \textit{RE}, and \textit{TLE} indicate wrong programs, such as wrong answer, runtime error, and time limit exceeded, respectively.}
    \label{fig:code_pairs}
\end{figure}

In the data collection phase, we collect the Python~3 programs submitted on AOJ and make code pairs consisting of WA and AC programs, as shown in Figure~\ref{fig:code_pairs}.
We target Python in this work because recent work on code-related tasks primarily targets Python.
In this work, WA includes wrong answers, logic errors, syntax errors, and time or memory limit exceeding.

We only use the code pairs whose edit distance is less than 100.
The code pairs with a large edit distance are more likely to be submitted to a wrong problem by mistake, or the changes in the code pair are not only the error fixes but also additional implementations or major changes in algorithms.
To analyze the frequent errors more accurately, we exclude code pairs that include such major changes.
We employ the \texttt{Levenshtein} library\footnote{\url{https://github.com/maxbachmann/Levenshtein}.} to calculate the edit distance of code pairs.

To analyze the user levels who submitted the program, we also collect metadata, including submission date, total submissions count at the time, attempts count to the problem, and whether it is the first acceptance.

\subsection{Code Normalization}
\label{sec:method:code-normalization}

\begin{table*}[t]
    \centering
    \begin{tabular}{llll} 
        \toprule
        Rule & Regular Expression & Rule Type & Description \\
        \midrule
        missing output & \multirow{3}{*}{\EscVerb{^print\\s\\(\\s.+?\\s\\)$|^print\\s\\(\\s\\)$}} & \textit{INSERT} & Add print statement \\ 
        needless output & & \textit{DELETE} & Delete print statement \\
        wrong output & & \textit{LINE-REPLACE} & Fix print statement \\
        \midrule
        wrong value & \EscVerb{^\\d+$|^\\d+\\.\\d*$|^\\.\\d+$} & \textit{TOKEN-REPLACE} & Fix numeric literal (int or float) \\
        \midrule
        wrong convert list & \EscVerb{\\slist\\s\\(\\s.+?\\s\\)|\\smap\\s\\(\\s\\w+\\s,\\s.+?\\s\\)} & \textit{WITHIN-REPLACE} & Fix conversion of list \\
        \bottomrule
    \end{tabular}
    \caption{Examples of classification rules.}
    \label{table:classification_rules}
\end{table*}

In the code normalization phase, to capture the abstract structure of the source code to classify the errors more accurately, we normalize the source code by removing unnecessary tokens.
While several prior works~\cite{yamaguchi2012ast, yoshizawa2018ast} used the abstract syntax tree (AST) to capture a more abstract structure than the normalized source code, the normalized source code can still capture superficial edits.

For normalization, we first tokenize the source code into a token sequence.
We employ the tokenizer tool provided by the CodeNet Project~\cite{puri2021codenet}.
By tokenization, unnecessary tokens are removed, such as comments, white lines, and white spaces.
In addition, to keep the structure of the source code, the tokens include \textit{control tokens}, which are not visible to the user, such as \texttt{NEWLINE}, \texttt{INDENT}, and \texttt{DEDENT}.

Variable and function names are often anonymized in the code normalization phase~\cite{osman2014classify, terada2021completion, shirafuji2023deduplicating}.
However, in this work, we omit the anonymization to detect the changes more accurately.
Anonymizing variable and function names lose significant information about variable or function names, and it will be difficult to detect the changes in the definition or invocation of a variable or function.

\begin{figure}[h]
    \centering
    \includegraphics[width=.8\linewidth]{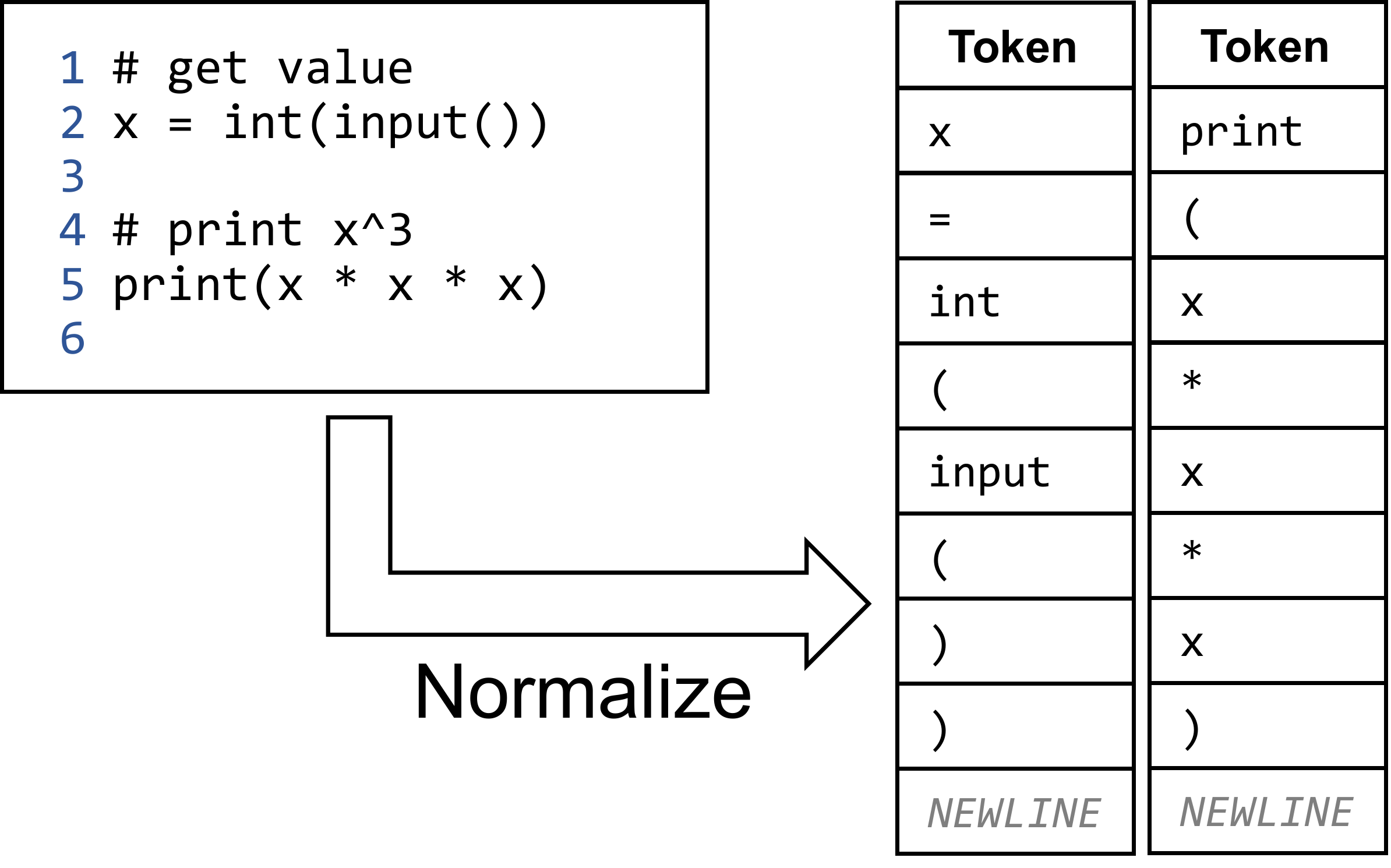}
    \caption{Example of code normalization. Comments, white lines, and white spaces are removed after tokenization. Variable and function names are retained (i.e., not anonymized). Control tokens (e.g., \texttt{NEWLINE}) are in italic.}
    \label{fig:normalization}
\end{figure}

As shown in Figure~\ref{fig:normalization}, the comments in the original program are removed, and white lines and white spaces are excluded in the token sequence.
The variable $x$ defined in the original program also appears as $x$ without anonymization in the token sequence.
In addition, while the token sequence contains control tokens (e.g., \texttt{NEWLINE}), these tokens are not visible to users.

\subsection{Changes Extraction}
\label{sec:method:changes-extraction}

In the changes extraction phase, we compare the WA and AC of normalized code pairs to extract the changed tokens.
This phase also involves labeling the changed type for each token.
We employ the \texttt{difflib} library\footnote{\url{https://docs.python.org/3/library/difflib.html}.} to extract the changes.

We first compare each line between the WA and AC.
If the lines are the same, all tokens of the lines are labeled \textit{EQUAL}.
Similarly, the added lines are labeled \textit{INSERT}, and the deleted lines are labeled \textit{DELETE}.
To identify which tokens are changed in the line, we further compare each token if the lines are different and likely to have been changed, and the added or deleted tokens are labeled \textit{REPLACE}.

\begin{figure}[h]
    \centering
    \includegraphics[width=\linewidth]{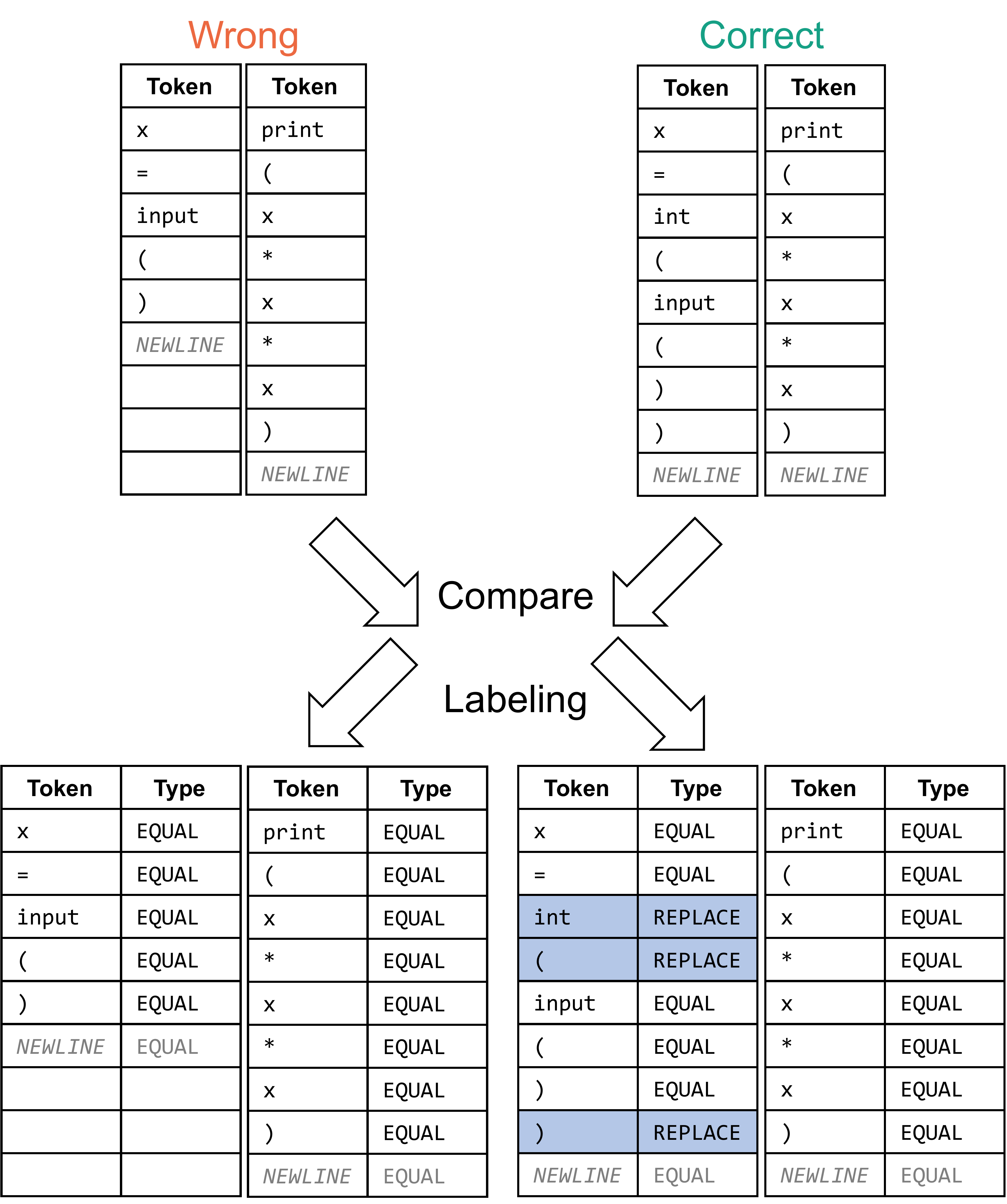}
    \caption{Example of changes extraction}
    \label{fig:changes_extraction}
\end{figure}

Figure~\ref{fig:changes_extraction} shows an example of the changes extraction.
This change includes an addition of the function invocation to convert a string to an integer. Since this addition is an inline replacement of tokens, these tokens are labeled \textit{REPLACE}.

\subsection{Error Classification}
\label{sec:method:error-classification}

In the error classification phase, we classify errors in the changed tokens by applying the 55 rules we defined.
We employ the \texttt{regex} library\footnote{\url{https://github.com/mrabarnett/mrab-regex}.} to use the regular expression.
This library supports additional features that the standard \texttt{re} library does not support, such as the negative lookbehind with the different lengths of alternatives.

The rules are divided into (1)~\textit{INSERT}, (2)~\textit{DELETE}, (3)~\textit{LINE-REPLACE}, (4)~\textit{WITHIN-REPLACE}, and (5)~\textit{TOKEN-REPLACE}.
The rules are used to detect what kind of lines of code are added, deleted, or replaced for what purpose.
\textit{INSERT} and \textit{DELETE} rules are applied to the lines labeled \textit{INSERT} or \textit{DELETE} in the changes extraction phase, \textit{LINE-REPLACE} and \textit{WITHIN-REPLACE} rules are applied to the lines containing \textit{REPLACE} tokens, and \textit{TOKEN-REPLACE} rule is applied to the tokens labeled \textit{REPLACE}.
Table~\ref{table:classification_rules} shows some examples of the rules and the corresponding regular expressions.

The difference between \textit{LINE-REPLACE} and \textit{WITHIN-REPLACE} is that when a regular expression matches a line, \textit{WITHIN-REPLACE} requires an additional check whether the \textit{REPLACE} token is contained within the matched range.
The defined regular expressions of \textit{LINE-REPLACE} rules contain~\textasciicircum\, at the beginning and \$ at the end to match the whole line.
Therefore, the regular expressions of \textit{LINE-REPLACE} rules already ensure that at least one \textit{REPLACE} token is contained in the matched range.
In contrast, as for \textit{WITHIN-REPLACE} rules, we further check whether the matched range contains at least one \textit{REPLACE} token because the regular expressions of \textit{WITHIN-REPLACE} rules can match the token sequence only containing \textit{EQUAL} tokens.

\begin{figure}[h]
    \centering
    \includegraphics[width=.95\linewidth]{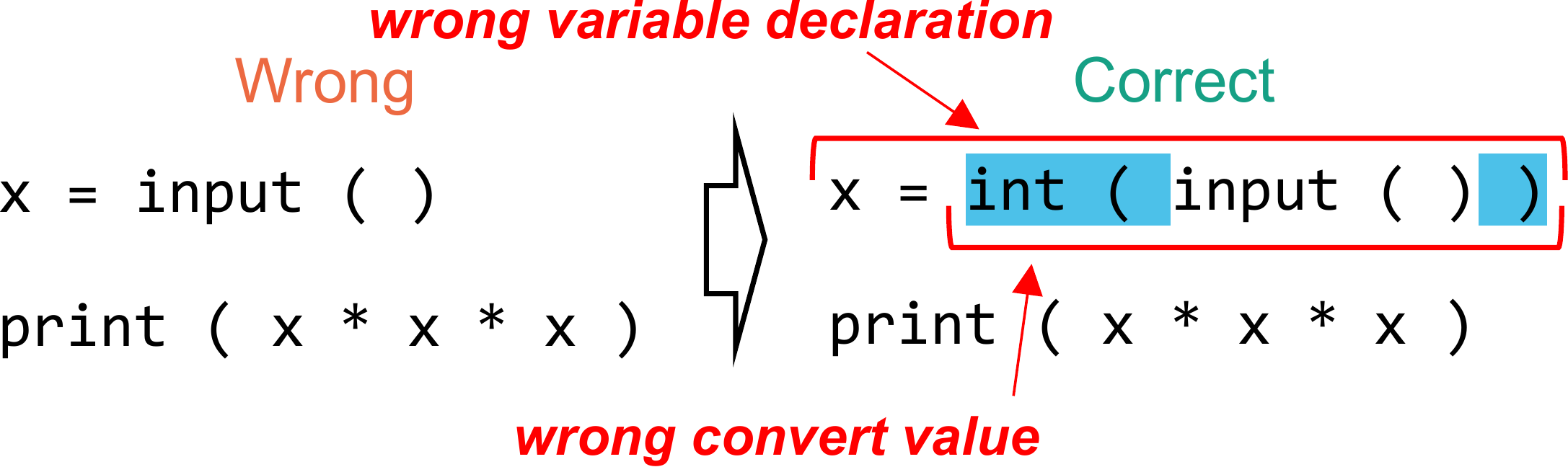}
    \caption{Example of error classification, which is classified as \textit{wrong variable declaration} (\textit{LINE-REPLACE}) and \textit{wrong convert value} (\textit{WITHIN-REPLACE}) because of missing int conversion of input value.}
    \label{fig:within_replace}
\end{figure}

Figure~\ref{fig:within_replace} shows the example of error classification for \textit{LINE-REPLACE} and \textit{WITHIN-REPLACE}.
In this example, the \textit{wrong variable declaration}, one of the \textit{LINE-REPLACE} rules, matches the whole line because it contains the = operator to represent the variable declaration.
In contrast, the \textit{wrong convert value}, one of the \textit{WITHIN-REPLACE} rules, matches only a few tokens directly associated with the \texttt{int()} conversion function.

We also define 21 summarized rules to analyze the summary of the errors.
Table~\ref{table:summarized_rules} shows the complete list of summarized rules and the original rules that belong to the summarized rule.
For example, the addition, deletion, or fix of \texttt{print()} invocation can be summarized as an \textit{output error}.
All the classified rules are summarized after the classification while the original classification results are also retained.

\begin{table}
    \centering
    \begin{tabular}{ll} 
        \toprule
        Summarized Rule & List of Classified Rules \\
        \midrule
        \multirow{3}{*}{output} & missing output \\
        & needless output \\
        & wrong output \\
        \midrule
        \multirow{3}{*}{input} & missing input \\
        & needless input \\
        & wrong input \\
        \midrule
        \multirow{3}{*}{convert variable} & wrong join list \\
        & wrong convert list \\
        & wrong convert value \\
        \midrule
        \multirow{3}{*}{other function invocation} & missing function invocation \\
        & needless function invocation \\
        & wrong function invocation \\
        \midrule
        \multirow{6}{*}{conditional statement} & missing if statement \\
        & needless if statement  \\
        & wrong if statement \\
        & missing else elif \\
        & needless else elif \\
        & wrong else elif \\
        \midrule
        \multirow{6}{*}{loop statement} & missing for statement \\
        & needless for statement \\
        & wrong for statement \\
        & missing while statement \\
        & needless while statement \\
        & wrong while statement \\
        \midrule
        for range & wrong range \\
        \midrule
        \multirow{3}{*}{break continue} & missing break continue \\
        & needless break continue \\
        & wrong break continue \\
        \midrule
        \multirow{2}{*}{literal} & wrong string \\
        & wrong value \\
        & wrong boolean value \\
        \midrule
        \multirow{3}{*}{import} & missing import \\
        & needless import \\
        & wrong import \\
        \midrule
        \multirow{3}{*}{variable declaration} & missing variable declaration \\
        & needless variable declaration \\
        & wrong variable declaration \\
        \midrule
        \multirow{6}{*}{function definition} & missing function definition \\
        & needless function definition \\
        & wrong function definition \\
        & missing return \\
        & needless return \\
        & wrong return \\
        \midrule
        \multirow{3}{*}{pass} & missing pass \\
        & needless pass \\
        & wrong pass \\
        \midrule
        comparison operator & wrong comparison operator \\
        \midrule
        logical operator & wrong logical operator \\
        \midrule
        arithmetic operator & wrong arithmetic operator \\
        \midrule
        unpack operator & wrong unpack operator \\
        \midrule
        \multirow{2}{*}{other operator} & wrong in operator \\
        & wrong assignment operator \\
        \midrule
        index & wrong list index \\
        \midrule
        list comprehension & wrong list comprehension \\
        \midrule
        indent & wrong indent \\
        \bottomrule
    \end{tabular}
    \caption{List of summarized rules.}
    \label{table:summarized_rules}
\end{table}

\subsection{Difference Analysis}
\label{sec:method:difference-analysis}

In the difference analysis phase, we define novice and expert programmers and analyze the difference in the frequent errors between the levels.
We define novice and expert programmers by the following criteria.
\begin{itemize}
    \item \textbf{Novices}: Only solved introductory problems, and the number of WA is more than five times the number of AC (i.e., $|\text{WA}| > |\text{AC}| \times 5$).
    \item \textbf{Experts}: Solved more than ten non-introductory problems.
\end{itemize}

Note that we set the number of errors classified in a code pair to up to one for each type of error.
Our early experiment has revealed that \textit{indent} fixes are sometimes made over many lines in a code pair, and \textit{function definition} fixes rarely appear many times in a code pair.
Therefore, it may cause a bias in the number of fixes.
In addition, even if the problem tended to include remarkable errors, there is a concern that the tremendous number of indent fixes would drown them out.
For these reasons, we exclude duplicated errors in this analysis phase.

We first use the chi-square test to examine the differences in the errors.
The chi-square test is denoted as Formula~\ref{formula:chi_square}, where $O$ is an observed frequency, and $E$ is an expected frequency.

\begin{equation}
    \chi^2=\sum_{k=1}^{n} \frac{(O_k - E_k)^2}{E_k}
    \label{formula:chi_square}
\end{equation}

Our null hypothesis is that \textit{``no difference of occurrence exists in the errors between novices and experts.''}
Therefore, our alternative hypothesis is that \textit{``there exist differences of occurrence in the errors between novices and experts.''}
We set the significance level $\alpha = 0.05$.
We reject the null hypothesis and accept the alternative hypothesis if the p-value of the $\chi^2$ is less than $\alpha$.

After the chi-square test, we use residual analysis to reveal the different errors.
The chi-square test can only test whether there exist differences.
Therefore, when we find a significant difference in the chi-square test, we perform a residual analysis to test for the specific error that makes the difference.

Standardized Pearson residual $r_{ij}$ is denoted as Formula~\ref{formula:residual_analysis}~\cite{agresti2013}, where $i$ is a row variable representing the classified errors, $j$ is a column variable representing novices or experts, $n_i$ is the row total number of frequencies, $n_j$ is the column total number of frequencies, and $N$ is the total number of frequencies.
\begin{equation}
    r_{ij} = \frac{O_{ij} - E_{ij}}{\sqrt{E_{ij} \cdot (1 - n_i/N) \cdot (1 - n_j/N)}}
    \label{formula:residual_analysis}
\end{equation}
The standardized Pearson residuals play the same role as z-scores.
Thus, we perform a two-tailed test and calculate the p-values for each cell.
In this analysis, our null hypothesis is that \textit{``no difference of occurrence exists in the error between novices and experts.''}
Therefore, our alternative hypothesis is that \textit{``there exists a difference of occurrence in the error between novices and experts.''}
When we find a significant difference in the residual analysis, we conclude that the error causes the difference between novices and experts.

\section{Results \& Discussion}
\label{sec:results}

We conduct two types of experiments.
The first experiment is to manually analyze the classification results and check the validity of the classification tool.
The second experiment is to test the differences in the errors for each introductory problem. 

In the experiments, we use \texttt{ITP1}\footnote{\url{https://onlinejudge.u-aizu.ac.jp/courses/lesson/2/ITP1/all}.}, which has 44 introductory problems available on AOJ, and the target source code is written in Python~3.

The total number of submitted programs on \texttt{ITP1} is 515,769, and the final number of constructed valid code pairs is 95,631.
Table~\ref{table:basic-dataset-statistics} and Table~\ref{table:additional-dataset-statistics} show the statistics of the constructed error-labeled dataset.

\begin{table}[h]
    \centering
    \begin{tabular}{lr} 
        \toprule
        Name & Value \\
        \midrule
        Source & AOJ ITP1 \\
        Language & Python~3 \\
        \#Problems & 44 \\
        \#Pairs & 95,631 \\
        \#Users & 10,361 \\
        Avg. \#Errors & 3.47 ($\pm$ 2.69) \\
        Avg. Char-based Edit Distance & 13.54 ($\pm$ 15.23) \\
        Avg. Token-based Edit Distance & 5.49 ($\pm$ 5.85) \\
        Avg. Char-based Similarity & 90.94\% ($\pm$ 11.13\%) \\
        Avg. Token-based Similarity & 89.44\% ($\pm$ 12.51\%) \\
        \bottomrule
    \end{tabular}
    \caption{Basic statistics of the error-labeled dataset.}
    \label{table:basic-dataset-statistics}
\end{table}

\begin{table}[h]
    \centering
    \begin{tabular}{lrr} 
        \toprule
        Name & WA Value & AC Value \\
        \midrule
        Avg. \#Lines & 8.02 ($\pm$ 7.78) & 8.04 ($\pm$ 7.76) \\
        Avg. \#Chars & 222.79 ($\pm$ 326.77) & 227.25 ($\pm$ 326.77)  \\
        Avg. \#Tokens & 80.76 ($\pm$ 105.67) & 82.14 ($\pm$ 105.73) \\
        Avg. Cyclomatic Complexity & 3.89 ($\pm$ 4.13) & 3.84 ($\pm$ 4.03) \\
        \bottomrule
    \end{tabular}
    \caption{Additional statistics of the error-labeled dataset. \textit{WA Value} and \textit{AC Value} indicate the value on average among the WA programs and AC programs, respectively.}
    \label{table:additional-dataset-statistics}
\end{table}

\subsection{Manual Evaluation}
\label{sec:results:manual}

As the evaluation of the classification tool, we randomly select 1,000 unique code pairs and manually validate the classified errors.
The selected 1,000 code pairs are classified as 3,704 errors (i.e., each code pair has 3.7 errors on average).
From our manual evaluation, 3,397 classified errors out of 3,704 are correct.
Therefore, the classification accuracy of the proposed tool is 91.71\%.

\subsection{Differences}
\label{sec:results:differences}

\begin{table*}[h]
    \centering
    \begin{tabular}{lp{5cm}rlrrr} 
        \toprule
        Problem & Problem Specification & P-value & Classified Rules & P-value & Novices & Experts \\
        \midrule
        \texttt{ITP1\_1\_A} & Print "Hello World". & $< .001$ & other function invocation & $< .001$ & 1.09\% & \textbf{3.85\%} \\
        \midrule
        \multirow{2}{*}{\texttt{ITP1\_1\_B}} & \multirow{2}{\linewidth}{Calculate the cube of a given integer.} & \multirow{2}{*}{.0390} & input & .0245 & \textbf{10.70\%} & 7.02\% \\
        & & & literal & .0339 & \textbf{14.11\%} & 10.17\% \\
        \midrule
        \multirow{3}{*}{\texttt{ITP1\_4\_B}} & \multirow{3}\linewidth{Calculate the area and circumference of a given circle radius.} & \multirow{3}{*}{$< .001$} & arithmetic operator & .0084 & \textbf{7.53\%} & 4.04\% \\
        & & & literal & .0025 & \textbf{22.34\%} & 15.41\% \\
        & & & convert variable & $< .001$ & 12.21\% & \textbf{24.55\%} \\
        \midrule
        \multirow{2}{*}{\texttt{ITP1\_4\_C}} & \multirow{2}{\linewidth}{Given two integers $a$ and $b$, and an operator $op$, print the value of $a\,op\,b$.} & \multirow{2}{*}{.0047} & arithmetic operator & .0261 & \textbf{19.59\%} & 13.28\% \\
        & & & comparison operator & $< .001$ & 2.03\% & \textbf{8.07\%} \\
        \midrule
        \multirow{2}{*}{\texttt{ITP1\_10\_A}} & \multirow{2}{\linewidth}{Calculate the Euclidean distance between two points $P1(x_1,y_1)$ and $P2(x_2,y_2)$.} & \multirow{2}{*}{$< .001$} & arithmetic operator & $< .001$ & \textbf{16.67\%} & 3.09\% \\
        & & & convert variable & .0400 & 30.95\% & \textbf{47.46\%} \\
        \bottomrule
    \end{tabular}
    \caption{Results of the analysis for each problem that showed significant differences. The third column indicates the p-value from the chi-square test, and the fifth column indicates from the residual analysis. The \textit{Novices} and \textit{Experts} indicate the ratio of the error frequency in the problem. The greater ratio is in \textbf{bold}.}
    \label{table:itp1_each_results}
\end{table*}

Table~\ref{table:itp1_each_results} shows the results of the analysis for each problem.
The table only includes problems and the summarized rule names that showed significant differences between novices and experts in the chi-square test and the residual analysis.
In the table, the third column represents the p-value from the chi-square test, and the fifth column represents the p-value from the residual analysis.
The sixth and seventh columns in the table represent the frequency ratio in the problem.
The greater ratio cell is marked as bold.

\subsubsection{ITP1\_1\_A}

In \texttt{ITP1\_1\_A}, we identify that experts tend to make errors of \textit{other function invocation} than novices.
The \textit{other functions} include functions that are not listed in other classification rules, e.g., string manipulation and mathematical functions.
In this problem, invoking functions other than \texttt{print()} is unnecessary because it only asks to output a specific string, ``Hello World''.
The reason for this result is that some experts try to solve the problem differently than usual by implementing their own print methods, such as calling system functions.
Moreover, since the problem is placed first and easy to access, some submissions contain completely unrelated programs, such as examining the execution environment of the OJ system, although the Terms of Use\footnote{\url{https://onlinejudge.u-aizu.ac.jp/term_of_use}.} prohibit such submissions.

\subsubsection{ITP1\_1\_B}

In \texttt{ITP1\_1\_B}, \textit{input error} and \textit{literal error} are more common among novices.
It results from trial and error by novices unfamiliar with the \texttt{input()} function because it is the first problem novices need to receive a value from standard input.
Also, instead of receiving input, some novices directly write a numeric literal given in the example case.
The submission will receive a verdict of the wrong answer since the value will be changed in the evaluation phase.

\subsubsection{ITP1\_4\_B}

In \texttt{ITP1\_4\_B}, \textit{arithmetic operator error} and \textit{literal error} are common among novices, and \textit{convert variable error} is common among experts.
The problem requires using $\pi$ to calculate the area of a given circle, and the error of the result digit is allowed up to 0.00001.

As for novices, they often define $\pi$ as a numeric literal rather than using the $\pi$ provided by the standard library (\texttt{math.pi}) due to their unfamiliarity with the standard library.
In many cases, the $\pi$ defined by novices does not have enough digits (e.g., $\pi = 3.14$), leading to incorrect results due to the large error in the calculation.

As for experts, because many of the problems on AOJ require conversions from string to an integer, some experts write an int conversion out of habit. However, the input constraint is a real number in this problem, and they need to write a float conversion instead.

\subsubsection{ITP1\_4\_C}

In \texttt{ITP1\_4\_C}, \textit{arithmetic operator error} is common among novices, and \textit{comparison operator error} is common among experts.

As for novices, the \textit{arithmetic operator error} is because of the difference between the / and // operators.
The difference between the two division operators in Python is that (1) / is to \textit{divide as a float} and retain the fractional part, and (2) // is to \textit{divide as an integer} and truncate the fractional part.
The / operator is typical to divide values in mathematics and many programming languages, and novices use it.
However, because the problem requires truncating any fractional part, using the~/ operator will result wrong.

As for experts, they misunderstand the terminal condition of the input, and it results in the high frequency of fixing the comparison operator.
The problem provides the terminal condition in the constraints section, but the condition can be estimated by referring to the input/output examples in the problem.
However, the actual condition is slightly different from the condition estimated from the examples, and it results in the fixing comparison operator.

\subsubsection{ITP1\_10\_A}

In \texttt{ITP1\_10\_A}, \textit{arithmetic operator error} is common among novices, and \textit{convert variable error} is common among experts.

As for novices, the arithmetic operator error is because of the difference between the \textasciicircum\, and ** operators.
The problem requires calculating Euclidean distance, which is denoted as $d=\sqrt{(x_1 - x_2)^2 + (y_1 - y_2)^2}$.
The square of $x$ can be denoted as \verb|x^2| in mathematics, and the novices write the same notation in Python code.
However, the \textasciicircum\, operator is a bit XOR operator in Python, and the ** operator is the correct operator.

The reason for the convert variable error among experts is the same as the \texttt{ITP1\_4\_B}; experts write an int conversion out of habit.

\section{Limitation}
\label{sec:limitation}

The set of problems used in this work, \texttt{ITP1}, is mainly solved by novice programmers as it is an introductory course.
There is a difference in the number of samples between novices and experts because novices submit more programs to introductory problems than experts, and experts make fewer errors in such introductory problems than novices.
However, for the same reason, the number of submissions will be larger for experts when we choose difficult problems.
Although AOJ is used by various levels of programmers, there are not so many problems equally solved by all levels of users because the users choose the problems that match their capability.
For future work, asking several levels of users to solve the same problem can make more significant differences to observe.

\section{Conclusion}
\label{sec:conclusion}

In this paper, we have proposed a rule-based error classification tool and analyzed its results.
The proposed tool can classify errors with 91.71\% accuracy according to our manual evaluation of the randomly selected classified errors.
The analyzed results showed that the errors made by novices are due to their fundamental lack of knowledge about the existence, usage, or specification of the functions.
On the other hand, the errors made by experts are due to misunderstandings caused by not reading the problems carefully.
For future work, we expect to be used for further code-related tasks, such as detecting potential errors and suggesting fixes for programmers using the proposed tool or the labeled dataset.

\bibliographystyle{ieeetr}
\bibliography{main}

\end{document}